\documentclass[conference,10pt]{IEEEtran}
\pdfoutput=1

\usepackage{listings}
\usepackage{color}

\usepackage[ruled, vlined, linesnumbered]{algorithm2e}
\usepackage{epsfig, subfigure}
\usepackage{amssymb, amsmath}
\usepackage{graphicx}
\usepackage{url}
\usepackage{subfigure}

\newcommand{\mian}[1]{{#1}}  

\begin{document}

\title{Optimizing the MapReduce Framework on \\ Intel Xeon Phi Coprocessor}

\author{\IEEEauthorblockN{$^1$Mian Lu\space$^2$Lei Zhang\space$^3$Huynh Phung Huynh\space$^4$Zhongliang Ong\space$^5$Yun Liang\space$^6$Bingsheng He\space$^7$Rick Siow Mong Goh\space$^8$Richard Huynh}
\IEEEauthorblockA{$^{1,3,4,7}$Institute of High Performance Computing, A*STAR \space\space $^{2,5}$Peking University \space\space  $^{6,8}$Nanyang Technological University\\
\{$^1$lum,$^3$huynhph,$^4$ongzl,$^7$gohsm\}@ihpc.a-star.edu.sg \space\{$^2$1000012927,$^5$ericlyun\}@pku.edu.cn \space$^6$bshe@ntu.edu.sg} }


\maketitle
\begin{abstract} 

With the ease-of-programming, flexibility and yet efficiency, MapReduce has become one of the most popular frameworks for building big-data applications. MapReduce was originally designed for distributed-computing, and has been extended to various architectures, e,g, multi-core CPUs, GPUs and FPGAs. In this work, we focus on optimizing the MapReduce framework on Xeon Phi, which is the latest product released by Intel based on the Many Integrated Core Architecture. To the best of our knowledge, this is the first work to optimize the MapReduce framework on the Xeon Phi.

In our work, we utilize advanced features of the Xeon Phi to achieve high performance. In order to take advantage of the SIMD vector processing units, we propose a vectorization friendly technique for the map phase to assist the auto-vectorization as well as develop SIMD hash computation algorithms. Furthermore, we utilize MIMD hyper-threading to pipeline the map and reduce to improve the resource utilization. We also eliminate multiple local arrays but use low cost atomic operations on the global array for some applications, which can improve the thread scalability and data locality due to the coherent L2 caches. Finally, for a given application, our framework can either automatically detect suitable techniques to apply or provide guideline for users at compilation time. We conduct comprehensive experiments to benchmark the Xeon Phi and compare our optimized MapReduce framework with a state-of-the-art multi-core based MapReduce framework (Phoenix++). By evaluating six real-world applications, the experimental results show that our optimized framework is 1.2X to 38X faster than Phoenix++ for various applications on the Xeon Phi.

\end{abstract}

\vspace{-0.3cm}
\section{Introduction}

Big data analytics has been identified as one of the most exciting
areas for both academia and industry. We are facing
the challenges at all levels ranging from sophisticated data mining algorithms
to high-performance computing techniques and systems to get the useful
data in time. The high-performance requirements come from the ever
growing data and time-consuming analytics processes.
High-performance system support for data analytics has become a
fruitful research area. Recently, we have
witnessed the success of various co-processors applying on data
analytics, such as graphics processors (GPUs)~\cite{Bingshengsigmod, gpustream} and FPGA~\cite{fpgamc}. In order to fully utilize the capability of those architectures, developers need to write co-processor specific programming languages (such as CUDA~\cite{cuda}, OpenCL~\cite{opencl} and Verilog~\cite{verilog}). This may affect the developer productivity, maintenance costs as well as the code portability and system scalability. Therefore, it is desirable to have high-performance accelerator systems with
compatible software development and maintenance with the CPU-based systems.

Recently, Intel released the long-awaited x86 accelerator
named Xeon Phi. It offers a much larger number of cores than conventional CPUs, while its architectural design
is based on x86. Particularly, an Intel Xeon Phi coprocessor 5110P
integrates 60 cores on a chip, with 4 hardware threads
per core. {The thread execution on the Xeon Phi does not suffer from the branch divergence. Furthermore, it highlights the 512-bit width vector processing units (VPUs)
for powerful SIMD processing. Besides, L2 caches are fully coherent through ring-based interconnection. It also provides low cost atomic operations. However, as designed as a coprocessor, the Xeon Phi has limited 8 GB main memory. }

While Xeon Phi has been just released, it has already demonstrated its promising adoptions. A number of studies
have demonstrated its performance advantage~\cite{xeonphi1, xeonphi2}.
The supercomputer STAMPEDE \cite{stampede} and Tianhe-2 \cite{tianhe2} also have equipped the Xeon Phi coprocessors to
unlock its hardware capability for scientific
computing. Instead of optimizing individual
applications like previous studies~\cite{xeonphi1, xeonphi2}, we
investigate a productivity programming framework to
facilitate users to implement big data analytics tasks correctly,
efficiently, and easily on Xeon Phi.

MapReduce~\cite{google} has become a popular programming framework
for big data analytics. It was originally proposed by Google for simplified parallel programming on a large number of
machines. Users only need to
define \textit{map} and \textit{reduce} functions according to their
application logics. The MapReduce runtime automatically distributes
and executes the task on multiple machines~\cite{google} or
multiple processors in a single machine~\cite{phoenix}, or
GPUs~\cite{mars}. Thus, this framework reduces the complexity of
parallel programming and the users only need to focus on the sequential
implementations of \textit{map} and \textit{reduce} functions.

A naive approach of developing a MapReduce framework on Xeon Phi is to adopt
Phoenix++ \cite{phoenix++} directly, which is the state-of-the-art MapReduce framework on multi-core CPUs. Since
Xeon Phi is also based on x86 architectures, Phoenix++ can run on
Xeon Phi without any changes. However, we find that Phoenix++ cannot fully utilize the hardware capability of Xeon Phi as Phoenix++ is not aware of the advanced hardware
features of Xeon Phi. First,
Phoenix++ pays little attention to utilize VPUs, which is
critical for the performance on the Xeon Phi~\cite{xeonphi1, xeonphi2}. Second, Phoenix++ has high memory access latency due to the
relatively small L2 cache (512 KB) per core. Third, the relatively small memory capacity (8 GB) may limit the thread scalability. Unawareness of
those hardware features results in significant performance loss, as
we demonstrated in Section~\ref{sec:experiment}.

To address the above-mentioned deficiency as well as fully utilize Xeon Phi hardware capabilities, we
develop \textbf{MRPhi}, the first MapReduce framework on Xeon Phi with following features.
\begin{itemize}
  \item  We implement the map phase in a vectorization friendly way to take advantage of the
SIMD VPUs.
  \item  SIMD hash computation algorithms are developed in order to benefit from the VPUs.
   \item Based on the MIMD hyperh-threading, we pipeline the map and reduce phases to improve the resource utilization.
  \item  We eliminate local containers by using low cost atomic operations on the global container in certain cases. This can address the thread scalability issue (due to the limited memory size) as well improve the cache efficiency (by utilizing the coherent L2 caches with ring interconnection).
  \item  Finally, these techniques are not applicable for all cases. For a given application, our framework is able to either automatically detect suitable techniques to apply or provide useful suggestions for users.
\end{itemize}




The rest of the paper is organized as follows.
We introduce the background in Section~\ref{sec:background}.
Section~\ref{sec:mrphi} gives detailed implementations. The
experimental results are presented in Section~\ref{sec:experiment}.
We conclude this paper in Section~\ref{sec:conclusion}.


\section{Background} \label{sec:background}
In this section, we first introduce the MapReduce framework and Xeon Phi coprocessor. Then we identify the challenges of developing the MapReduce framework on the Xeon Phi.
\vspace{-0.2cm}

\subsection{MapReduce Framework}
\vspace{-0.1cm}
MapReduce is a popular framework for simplified parallel programming. We briefly introduce the MapReduce framework.

\textbf{Programming model.} The input of a MapReduce job is specified by users, usually it is an array. The output is a set of \textit{key-value} pairs. A user specifies a MapReduce job mainly by two functions, which are \textit{map} and \textit{reduce}. With the user-defined functions, a MapReduce framework first applies the map function to every element in the input array and generates a set of intermediate key-value pairs (\textit{map phase}). After the map phase, the reduce function is applied to all intermediate pairs with the same key and generates another set of result key-value pairs (\textit{reduce phase}). Finally, the result key-value pairs are ordered (optional) and then output. The detailed programming model is presented in the original MapReduce paper \cite{google}.



\textbf{MapReduce workflow.} The MapReduce framework is originally designed for distributed computing~\cite{google}. Later, it is extended to other architectures such as multi-core CPUs~\cite{phoenix, phoenix++, tiled, metis}, GPUs~\cite{mars, multigpu, mapcg}, the coupled CPU-GPU architecture~\cite{apu}, FPGA~\cite{fpga} and Cell processors~\cite{cell}. These different MapReduce frameworks share the common basic workflow, but differ in detailed implementation and optimization.

\begin{figure}[ht]
\centering
\includegraphics[scale=0.30]{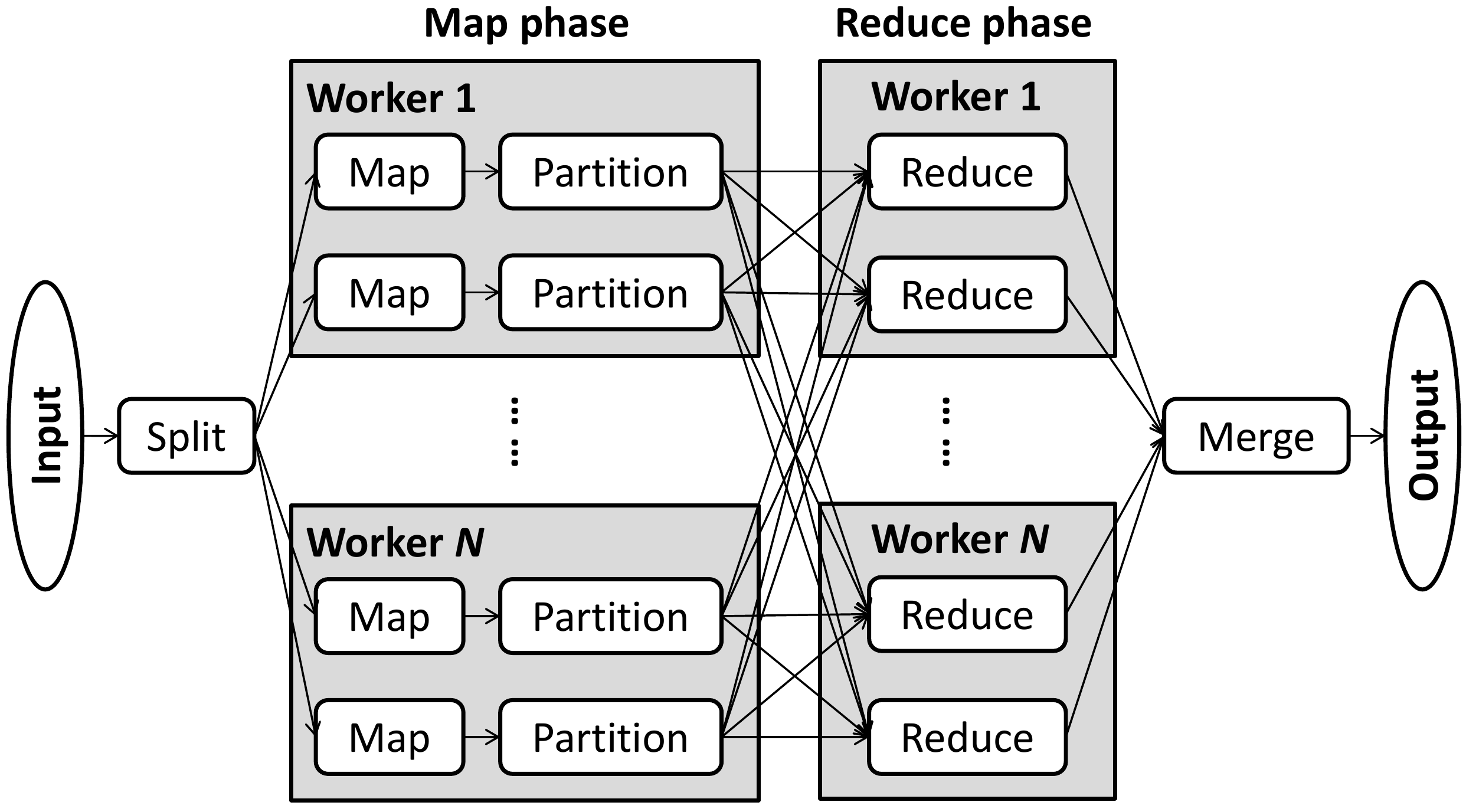}
\vspace*{-.3cm}
\caption{The basic workflow of a MapReduce framework.}
\label{fig:basic_workflow} \vspace*{-.3cm}
\end{figure}

Figure~\ref{fig:basic_workflow} illustrates the basic workflow of a MapReduce framework. At the beginning, a \textit{split} function (either defined by the system or users) divides the input data across \textit{workers}. On multi-core CPUs, a worker is handled by one thread. A worker usually needs to process multiple input elements. Thus the \textit{map} function is applied to the input elements one by one. Such a call of the map function for an input element is called a \textit{map operation}. Each map operation produces a set of intermediate key-value pairs. Then a \textit{partition} function is applied to these key-value pairs according to the key. Then in the reduce phase, each \textit{reduce operation} applies the reduce function to a set of intermediate pairs with the same key. Finally the results from multiple reduce workers are merged and output.

\vspace{-0.1cm}
\subsection{Intel Xeon Phi Coprocessor} \label{sec:xeonphi}
\vspace{-0.1cm}
Intel Xeon Phi coprocessor is recently released in November 2012. The Xeon Phi is based on Intel Many Integrated Core Architecture. The current released product is 5110P. Overall, the Xeon Phi 5110P integrates 60 x86 cores on the same chip. Each core has the frequency of 1.05 GHz and supports 4 hardware threads. The memory hierarchy of Xeon Phi is similar to a conventional multi-core CPU. The memory refers to the main memory on the Xeon Phi, which is shared and accessible for all cores. The main memory size is 8 GB. Then each core does not have its own local memory, but has local L1 and L2 caches. The caches are managed by the underlying system as a conventional CPU. The L2 cache size is 512 KB per core. Additionally, on each core, there are 32 512-bit vector registers.



Xeon Phi has been used to accelerate linear algebra~\cite{xeonphi2} and molecular dynamics~\cite{xeonphi1} for high performance. These studies show good performance potential of Xeon Phi when the important hardware capabilities are utilized, e.g., 512-bit SIMD vector processing units. Furthermore, Xeon Phi coprocessors also start to play an important role for supercomputers, such as STAMPEDE \cite{stampede} and Tianhe-2 \cite{tianhe2}. In the following, we briefly introduce the major features of the Xeon Phi.

\textbf{512-bit vector processing units (VPUs).} Xeon Phi features with wide 512-bit VPUs on each core. It doubles the vector width compared with the latest Intel Xeon CPU. Furthermore, it provides new SIMD primitives, such as scatter/gather. Therefore, utilizing VPUs effectively is the key to deliver high performance. The VPUs can be either exploited by manually implementations with SIMD instructions or \textit{auto-vectorization} by the Intel compiler. The auto-vectorization tries to identify loops that can be vectorized to use SIMD VPUs at compilation time.

\textbf{MIMD Massive thread parallelism.} Each core of the Xeon Phi supports up to 4 hardware hyper-threads. Thus, there are 240 threads in total. 
The MIMD (Multiple Instruction, Multiple Data) thread execution allows different threads execute different instructions at any time. Thus, we can assign different workloads to different threads to improve the hardware resource utilization.

\textbf{Coherent L2 caches with ring interconnection.} The interconnection on the Xeon Phi employs a ring architecture. All L2 caches are coherent through the ring interconnection. This design is able to improve the cache efficiency. Specifically, when a cache miss occurs on a core ($C_0$), an address request is sent to the address ring. If the data is found in another core's L2 cache, the data will be forwarded to the original core $C_0$ by the data block ring. This way, it avoids the expensive memory access.

\textbf{Low cost atomic operations.} Atomic data types are well supported on the Xeon Phi. Compared with native data types, the operations on atomic data types do not have significant overhead. Therefore, it is reasonable to exploit the use of atomic operations when designing a parallel algorithm.

However, the memory size of the Xeon Phi is fixed and small compared with traditional main memory, which is only 8 GB. This may become a bottleneck when designing efficient algorithms. We demonstrate such an issue for particular applications and use low cost atomic operations to address it in Section \ref{sec:global}.

There are two modes to execute a program using the Xeon Phi, which are \textit{offloading} and \textit{native} execution. In this study, we use the native execution. We will develop the offloading version in the future work. A native program entirely runs on the Xeon Phi. There is no communication with the host. Additionally, Xeon Phi is compatible with traditional parallel programming languages, such as OpenMP, pthread, OpenCL, MPI and so on.

\subsection{Challenges of a Shared Memory MapReduce Framework on Xeon Phi} \label{sec:issue}
\vspace{-0.1cm}
State-of-the-art shared memory MapReduce frameworks on multi-core (such as Phoenix++ \cite{phoenix++}) are designed to have flexible intermediate key-value storage container targeting different kinds of workloads and effective combiner implementation that perform reduce function immediately on the map results locally in each core~\cite{tiled}. These techniques reduce memory storage requirement and traffic as well as increase data locality. However, we have identified three major performance issues of Phoenix++ when porting it onto Xeon Phi.

\begin{itemize}

\item{\textit{Poor VPU usage}. Phoenix++ takes little advantage of the VPUs on the Xeon Phi. The compiler is unable to vectorize the code effectively. This suggests that we should rewrite the code in a suitable way to assist the auto-vectorization.}

\item{\textit{High memory latency}. Container (hash table or arrays) building has a large number of random memory accesses. As the local L2 cache per core on the Xeon Phi is small, this results in high cache misses and memory latency.}

\item{\textit{Small memory}. Due to the limited memory (8 GB) on the Xeon Phi, we find that Phoenix++ cannot handle the array container efficiently when the array is large.}
\end{itemize}

As a result, running Phoenix++ directly on the Xeon Phi does not give good performance. In our framework, we propose various techniques to address these performance issues.

\section{Optimized MapReduce Framework on Xeon Phi} \label{sec:mrphi}

In this section, we present our proposed MapReduce framework \textit{MRPhi}, which is specially optimized for the Xeon Phi.
\vspace{-0.2cm}
\subsection{Overview}
\vspace{-0.1cm}
MRPhi adopts state-of-the-art techniques from the shared memory MapReduce framework as well as specific optimizations for the Xeon Phi coprocessor. Overall, we adopt the similar design as Phoenix++ \cite{phoenix++} to implement the basic MapReduce workflow as shown in Figure \ref{fig:basic_workflow}. There are two major techniques from Phoenix++ adopted in our framework, which are efficient combiners and different container structures. We briefly introduce the two techniques and refer readers to the original Phoenix++ paper \cite{phoenix++} for more details.
\begin{itemize}
\item{\textit{Efficient combiners}. Each map worker maintains a local container. When an intermediate key-value pair is generated by a map function, the reduce operator is immediately applied to that pair based on the local container. This process is performed using a \textit{combiner}. Therefore, a map operation essentially consists of two parts, which are the computation defined in the map function and combiner execution. After that, the partition is applied to each local container and multiple local containers are merged to a global container in the reduce phase. Our MRPhi adopts this similar design but with particular improvement (Section \ref{sec:global})}. Note that both local and global containers are stored in the main memory of Xeon Phi.
\item{\textit{Hash table and array containers}. MRPhi supports two data structures for containers, which are hash table and array. The array container is efficient when the keys are integers and in a fixed range.}
\end{itemize}

\begin{figure}[ht]
\centering
\includegraphics[scale=0.52]{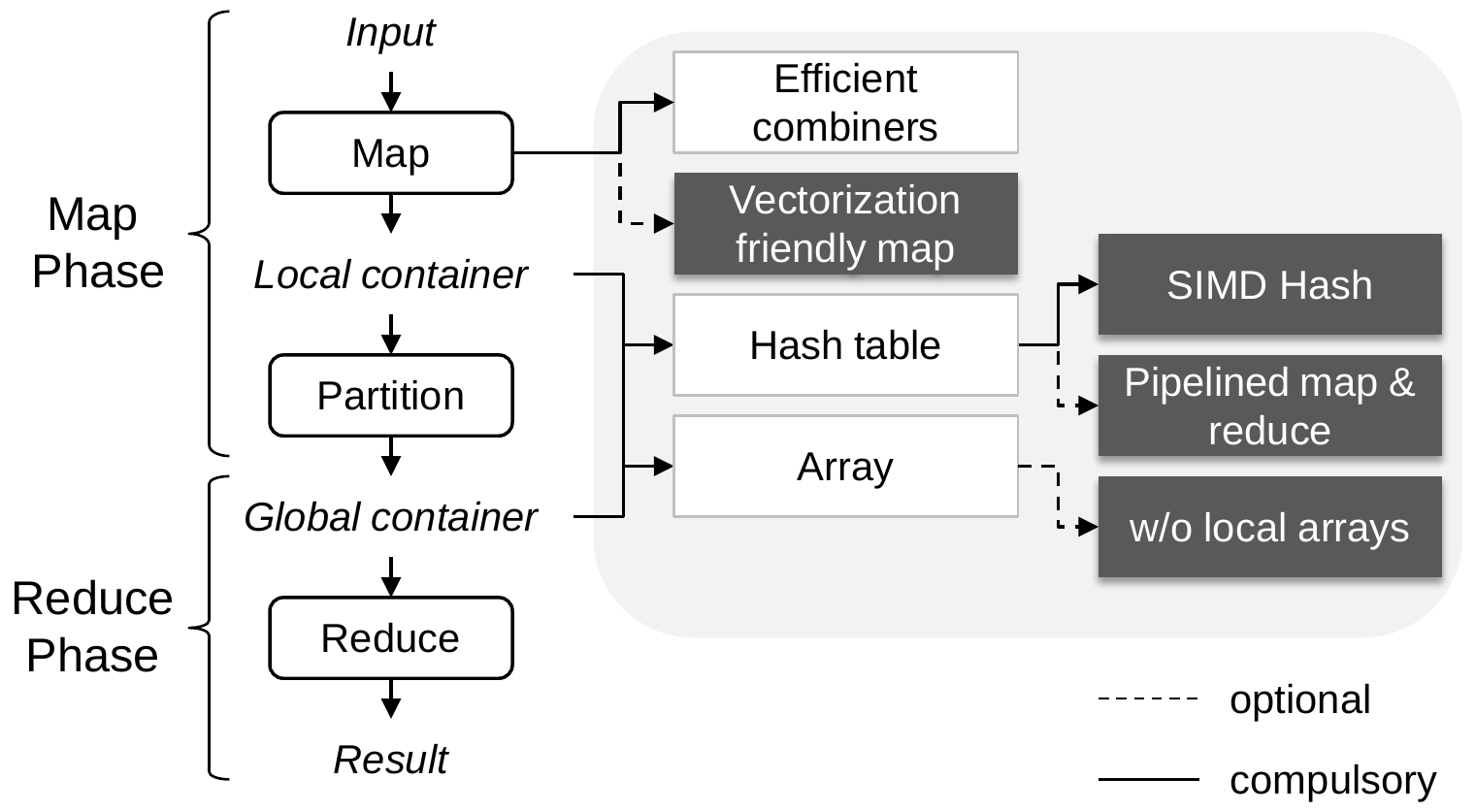}
\vspace*{-.3cm}
\caption{Proposed techniques (in dark box) and their applicability in MRPhi.}
\label{fig:opt_tech} \vspace*{-.3cm}
\end{figure}

More importantly, we propose four optimization techniques specific for Xeon Phi as shown in Figure \ref{fig:opt_tech}. The left part of the figure summarizes the flow of MapReduce framework while the right part shows the optimization techniques applied correspondingly. On the right part, the white boxes are the adopted combiners and containers from Phoenix++ while the dark boxes are our proposed optimization techniques.

\begin{itemize}
  \item {\textbf{Vectorization friendly map phase.} MRPhi implements the map phase in a vectorization friendly way, which clears the dependency among map operations. By doing this, the Intel compiler can automatically vertorize multiple map operations to take advantage of VPUs successfully.}
  \item {\textbf{SIMD parallelism for hash computation.} Hash computation can be implemented employing SIMD parallelism. We implement it using SIMD instructions.}
  \item {\textbf{Pipelined execution for map and reduce phases.} In general, the user-defined map function contains heavy computation workload, while the reduce function has many memory accesses \cite{phoenix++}. In order to better utilize the hardware resource with hyperthreading, we pipeline the map and reduce phases based on the MIMD thread execution.}
  \item {\textbf{Eliminating local arrays.} For the array container, if the array is large, it introduces a number of performance issues due to local arrays adopted. We address these issues by eliminating local arrays but employing low overhead atomic operations on the global array. It also can improve the cache efficiency because of the coherent L2 caches with ring interconnection.}
\end{itemize}

Note that these techniques are not applicable for all cases and may introduce overhead. Our framework can either automatically detect whether a specific technique is applicable or provides helpful suggestions to users at compilation time.
\vspace{-0.2cm}
\subsection{Vectorization Friendly Map Phase}
\vspace{-0.1cm}
Utilizing VPUs is critical to high performance on the Xeon Phi. For the MapReduce framework itself, except the hash computation (presented in Section \ref{sec:hash}), there is little chance to employ SIMD instructions. However, the user-defined map function either containing loops or not is potential for auto-vectorization. For the case of containing loops, we leave the compiler to identify the vectorization within the map function. Our main focus is on the vectorization challenge for the case of not containing loops in the user-define map function.

Recall that in the map phase, each thread (or map worker) processes multiple map operations. We use directives to guide the compiler to vectorize multiple map operations. Listing \ref{list:vec} shows the basic idea. \textit{emit\_intermediate} is a system-defined function to perform combiners. The \textit{\#pragma ivdep} (line 3) tells the compiler to vectorize this for-loop if there is no dependency.

\lstset{basicstyle=\footnotesize\ttfamily, frame=bt,caption=Vectorization for multiple map operations, captionpos=b,numbers=left,xrightmargin=9pt,xleftmargin=20pt,breaklines=true,tabsize=2, label={list:vec}}
\begin{lstlisting}
//N: the number of map operations in the worker
//elems: the input array
#pragma ivdep
for(i = 0; i < N; i++) {
    //the inlined map function
    map(data_t elems[i]) {
        ... //some computation
        emit_intermediate(key, value);
        ...
    }
}
\end{lstlisting} \vspace{-0.2cm}

\begin{figure}
\subfigure[The original design of map operations
]{\includegraphics[scale=0.28]{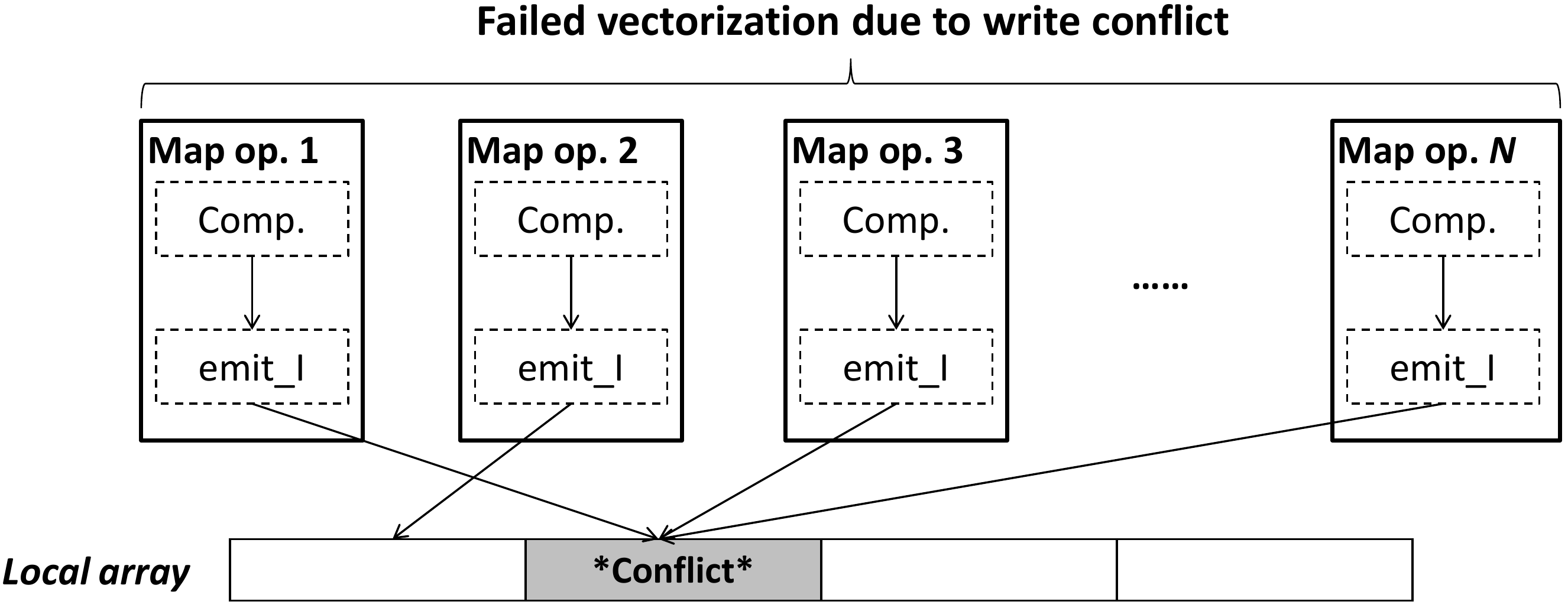} \label{fig:no_vec_friendly}}
\vfill \subfigure[Vectoriation friendly map]{\includegraphics[scale=0.28]{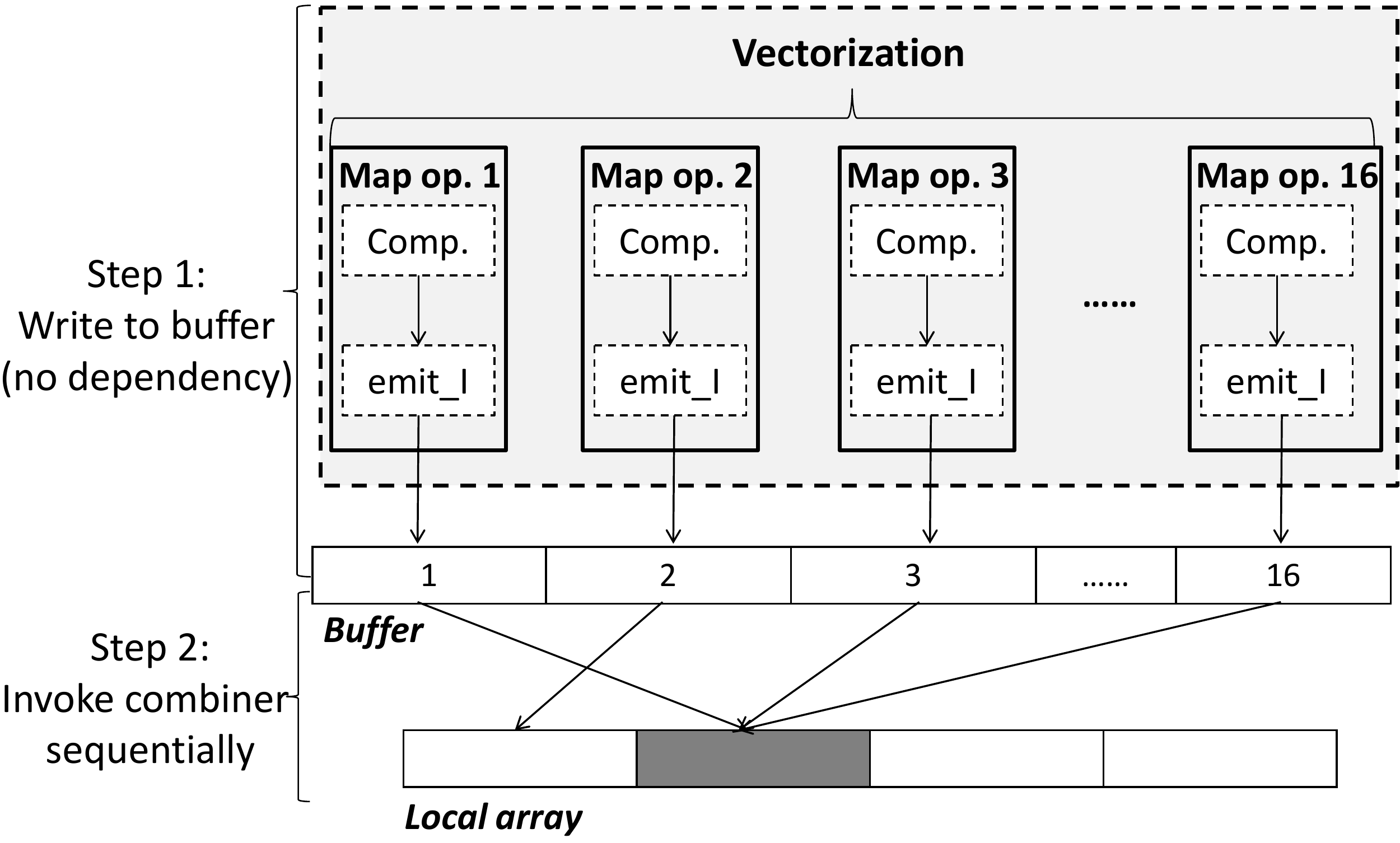} \label{fig:vectorization_friendly_b}}\vspace{-0.3cm}
\caption{Comparison of different designs for the map phase.}\vspace{-0.6cm}
\label{fig:vectorization_friendly}
\end{figure}

This auto-vectorization can be effective if there is no dependency among map operations (from line 6 to 10 in Listing \ref{list:vec}). However, in Phoenix++, if multiple \textit{emit\_intermediate} operations are performed concurrently, the execution will cause the conflict on a local container. This conflict exists for either array or hash table containers. Figure \ref{fig:no_vec_friendly} illustrates an example that map operations failed to be vectorized due to the dependency among \textit{emit\_intermediate} for an array container.


We propose a vectorization friendly technique to address this issue. Instead of performing the combiner for each intermediate pair generated by \textit{emit\_intermediate} immediately, we buffer a number of pairs. Writing to the buffer is independent for each map operation. When the buffer is full, we call the combiner for those pairs sequentially. Figure~\ref{fig:vectorization_friendly_b} demonstrates this vectorization friendly map. Our vectorization friendly map clears the dependency among map operations and thus auto-vectorization by the compiler is possible.

The vectorization friendly map is useful when the multiple map operations can be vectorized (no loop in the map function). On the other hand, this technique introduces overhead due to the temporary buffer. Therefore, if this technique is enabled but the map operations cannot be vectorized, it will hurt the performance. Since we rely on the compiler to enable vectorization, the framework itself does not know until the compilation for a specific application. If the map operations can be vectorized based on the printout from Intel compiler about vectorization eligibility, then it is worthwhile to adopt this technique. Due to the clear output by the compiler and our clean design of the interface, turning on or off this technique essentially is very easy.
\subsection{SIMD Parallelism for Hash Computation} \label{sec:hash}
Hash computation is a key component in the MapReduce framework as well as fundamental for many other data-intensive applications, such as database and encryption systems. We observe that the auto-vectorization often fails due to the complex logic for hash computation. Thus, we choose to manually implement the hash computation using SIMD instructions.


SIMD hash computation for native data types is straightforward. The same procedure is applied to different input elements, which fully employs the SIMD feature. However, it is challenging to process variable-sized data types, such as text strings. Overall, various hash functions for strings, such as FNV \cite{fnv} and djb2 \cite{djb2}, have the similar workflow, which processes characters one by one as shown in Listing \ref{list:hash}. As a result, the workload of the hash computation for a given string depends on its length. The challenge is how to handle variable lengths yet be efficient. Furthermore, SIMD instructions only can be applied to special VPU vector registers. Thus how to pack data from memory to these vectors efficiently is another challenging problem.

\begin{lstlisting}[language=C,frame=bt,caption=The workflow of hash functions for strings,captionpos=b,numbers=left,xrightmargin=9pt,xleftmargin=20pt,breaklines=true,tabsize=2,label={list:hash}]
int hash(char* str) {
    v = ...//initialization
    while (*str) {
        v = func(v, *str++);
    }
    return v;
}
\end{lstlisting}\vspace{-0.2cm}

We propose two SIMD hash computation algorithms. The first one is easy to implement and fully takes advantage of SIMD scatter/gather, but may lead to low SIMD hardware utilization. The second one improves  the SIMD hardware utilization but at expense of high control flow overhead. In the following, we name these two implementations as \textit{SIMDH-Padding} and \textit{SIMDH-Stream}.

\subsubsection{SIMDH-Padding}
It contains multiple rounds and each round processes characters from 16 consecutive strings in parallel. The intuition is within each round, we treat 16 strings as equal-length strings with the length $L_r$. $L_r$ is equal to the number of characters in the longest string among these 16 strings. Then if a string is shorter than $L_{r}$, we pad this string with empty characters. Note that this padding in fact is implemented using masks for efficiency.

SIMDH-Padding has significant low control overhead due to its simplicity. It takes full advantage of SIMD instructions and also able to utilize the SIMD gather for data packing. However, it underutilizes the computation resource due to the padding of empty characters.

\subsubsection{SIMDH-Stream}
\begin{figure}[ht]
\centering
\includegraphics[scale=0.47]{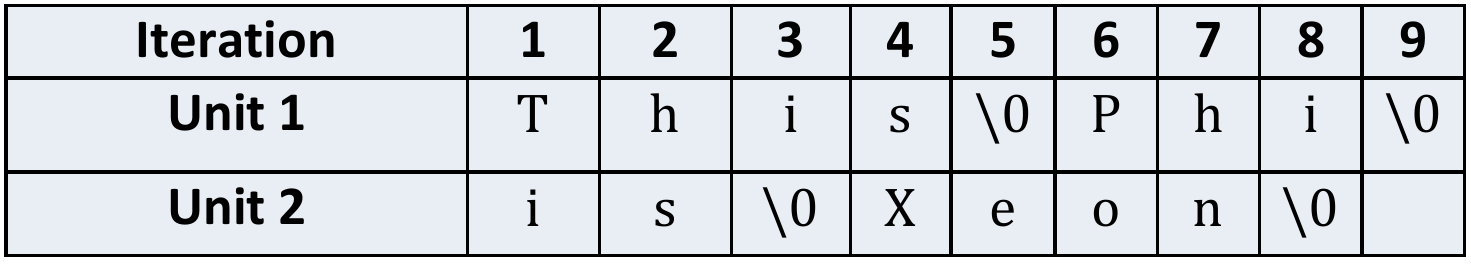}
\vspace*{-.0cm}
\caption{The workflow of SIMDH-Stream.}
\label{fig:simd2} \vspace*{-.4cm}
\end{figure}

This algorithm does not divide strings to groups for different rounds. Instead, we continuously feed the SIMD units with strings. We treat the input strings as a stream. In Figure~\ref{fig:simd2}, the zero ($\setminus0$) denotes the end of a string. Suppose we process two strings (16 strings in practice) in parallel using two SIMD units. The input array contains four strings: ``This'', ``is'', ``Xeon'', ``Phi''. First, we start to process words ``This'' and ``is''. In the third iteration, the word ``is'' in the second unit is finished. Then we immediately pack the first character ``X'' from next word ``Xeon'' into the second unit and continue the process. Then in the 5th iteration, the word ``This'' is finished in the first unit. We pack the first character ``P'' from the word ``Phi'' into the first unit.

For the data packing, we adopt a prefix-sum based method as well as utilizing SIMD primitives. First, we use one SIMD instruction to know which characters in the vector are zeros and store the result in a mask vector. Then we perform prefix-sum on the mask vector to obtain the index of next strings to process. Finally, the SIMD gather is used to collect the characters. In order to allow fast prefix-sum computation we use a lookup table. Since each vector contains 16 32-bit elements, we store the prefix-sum result for 8 elements instead of 16 elements for space efficiency. As a result, a lookup table of $2^8$ entries with the size of around 2 KB is used, which can fit into the cache.

Compared with SIMDH-Padding, SIMDH-Stream introduces more complex control flow for data packing. SIMDH-Padding simply checks whether the 16 packed characters are all equal to zero in each iteration. If it is true, then we pack all first characters of the next 16 strings to the vector. These actions can be finished by two SIMD instructions (one comparison and one gather). However, in SIMDH-Stream, we check characters in the vector one by one for each iteration. If a character is zero, then we load the first character of the next unprocessed string to the vector. There is no direct SIMD instructions for this process. However, SIMDH-Stream does not waste computing resources on empty characters. We evaluate these two algorithms in Section \ref{sec:exp_study}.

\vspace{-0.1cm}
\subsection{Pipelined Execution for Map and Reduce Phases} \label{sec:pipeline}\vspace{-0.1cm}

Pipelined map and reduce has been adopted in the MapReduce framework for distributed computing to improve the performance \cite{online}. We propose to pipeline map and reduce phases on the Xeon Phi based on the MIMD thread execution. The motivation is that the map function defined by users usually performs heavy computation. But the reduce phase contains many memory accesses in which the major work is to construct the global container. We pipeline the computation-intensive map and memory-intensive reduce to improve the overall hardware resource utilization. This technique is more effective for the hash table container. Because the time of reduce phase when using the array container is usually too short to take advantage of pipelining.

\begin{figure}[ht]
\centering
\includegraphics[scale=0.32]{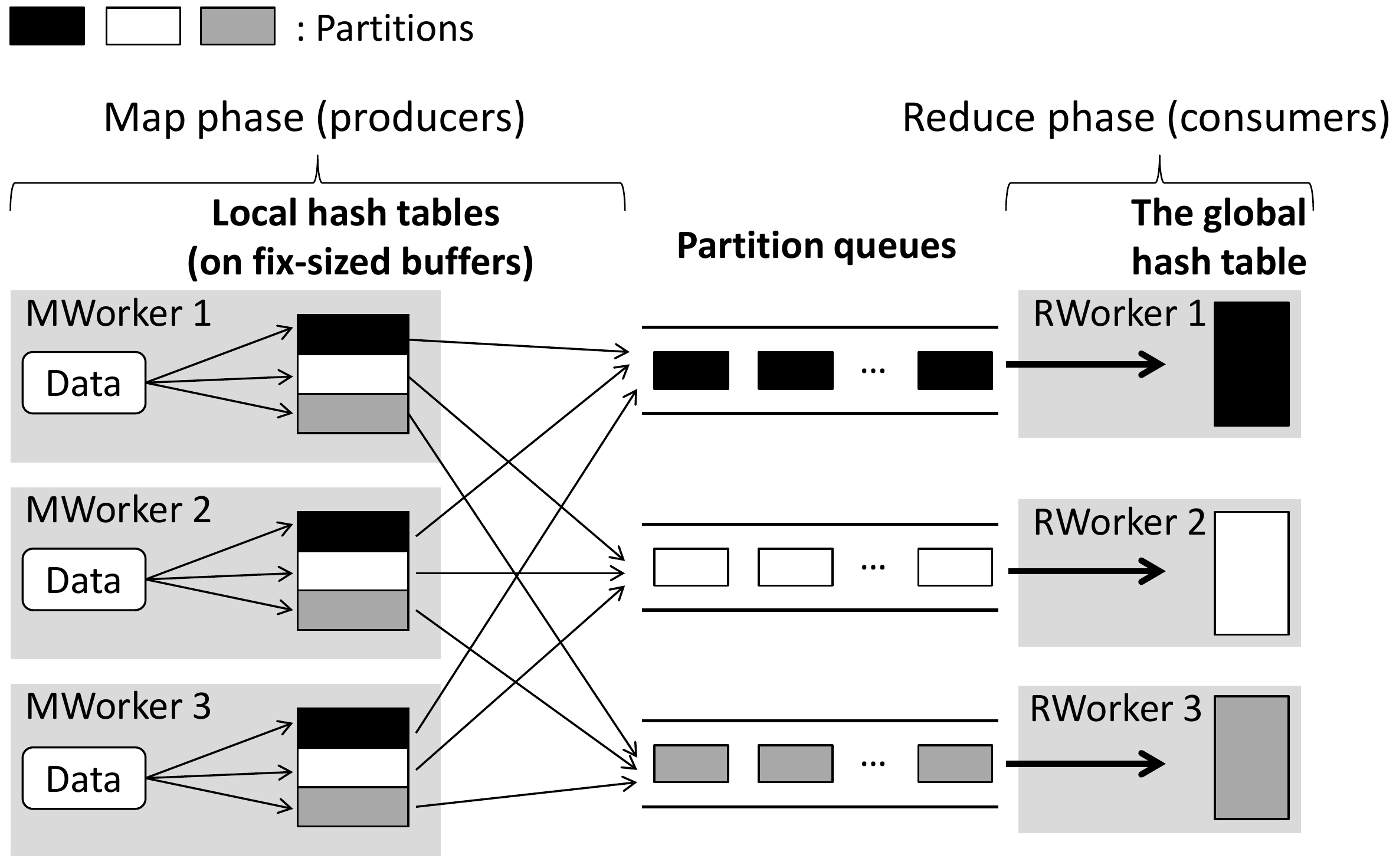}
\vspace*{-.0cm}
\caption{The producer-consumer model for the pipelined map and reduce phases.}
\label{fig:pipeline} \vspace*{-.3cm}
\end{figure}

We design an efficient producer-consumer model to pipeline the map and reduce phases. There are three major data structures, which are local hash tables, a global hash table, and partition queues. Specifically, each map worker still has a local hash table. However, the local table works on a pre-allocated fix-sized small buffer, e.g., smaller than the L2 cache. This is to improve the data locality of local hash table building. There are First-In-First-Out queues for different partitions. The \textit{push} and \textit{pop} operations on the queues are performed by the map and reduce workers, respectively. Suppose there are $N_r$ reduce workers, then there are $N_r$ queues.

Figure \ref{fig:pipeline} illustrates an example of three producers and three consumers. There are concurrent map workers (\textit{MWorker}) and reduce workers (\textit{RWorker}) on each core. It works in this way. Each map worker processes the data and builds its own local hash table. When the local hash table size exceeds the buffer size, it partitions the current table and sends partition $i$ to queue $i$. After that, the local buffer is empty, and the map worker starts to build a new local table. On the other hand, each reduce worker corresponds to one queue. As long as the queue is non-empty, it fetches the partitions and builds the global hash table. Since data are partitioned, there is no conflicts among reduce workers.


If the final global hash table is very small, e.g., smaller than the L2 cache, the non-pipelined model will be more efficient. The major reason is the reduce phase will be too short to take advantage of pipelining because of the small hash table. On the other hand, the pipelined model introduces storage overhead. Our producer-consumer model is adaptive to this case. Recall that we allocate a fix-sized buffer (smaller than the L2 cache) for the local hash table. If the final hash table is smaller than this buffer, no data will be fed to the reduce worker (the consumer) until the map phase is finished. This way, our pipelined model essentially degrades to a non-pipelined model as we expected.
\vspace{-0.1cm}
\subsection{Eliminating Local Arrays} \label{sec:global}\vspace{-0.1cm}

Recall that in order to support efficient combiners, Phoenix++ uses a local container for each worker in the map phase. Then the local containers are merged in the reduce phase for the final result. This design is efficient when the container size is small. However, it will introduce performance issues when the container becomes large. An alternative is to eliminate local containers and directly update on the global container with low-cost atomic operations for combiners when the container size is large.

This technique is applied to the array container. Because the atomic data types only support basic arithmetics while the hash table usually requires more complex data types and operations, such as text strings and memory allocation. Based on the low overhead of atomic data types on the Xeon Phi, using the global array directly is more efficient when the array becomes large. There are two major advantages.

\textbf{Thread scalability.} Due to the relatively small memory size on the Xeon Phi (8 GB), the thread scalability can be limited when using local arrays. Note that the local array is allocated in the memory of Xeon Phi. Suppose the local array size is $L$ bytes, the available memory is $M$ bytes, then the maximum number of concurrent threads for the map phase is $\lfloor{\frac{M}{L}\rfloor}$. As an extreme example of using Bloom filter in bioinformatics \cite{bloomfilter} (evaluated in Section \ref{sec:experiment}), if the whole human genome is used, the local array size is around 3.7 GB. In such a case, only two threads can be used on the Xeon Phi employing local arrays.

\textbf{Cache efficiency.} When the array is small to fit into the L2 cache, using local arrays has good cache efficiency. However, when the array becomes large, random memory accesses on local arrays cause poor data locality. Eliminating the local arrays but using the global array directly for combiners is able to improve the cache efficiency by the ring interconnection for L2 caches on the Xeon Phi. Specifically, when using local arrays, every L2 local cache miss should cause a memory access. On the contrary, when using the global array directly, the global array is shared across multiple cores. When a L2 cache miss occurs on one core, the data may be copied from another core's L2 cache to avoid the expensive memory access. This takes advantage of the ring-based interconnection architecture on the Xeon Phi.



Whether eliminating local arrays is decided by our framework automatically. Specifically, we mainly consider the cache efficiency. If the size of each local array is smaller than the L2 cache, then we keep these small local arrays as Phoenix++ does. Otherwise our framework will eliminate the local arrays.

\vspace{-0.1cm}
\section{Experimental Evaluation} \label{sec:experiment}
\vspace{-0.1cm}


\textbf{Hardware setup.} We conduct our experiments on an Intel Xeon Phi coprocessor 5110P. The hardware specification has been summarized in Section \ref{sec:xeonphi} and the Intel compiler is used.

\vspace{-0.2cm}
\begin{table} [h]
  \caption{Benchmark applications.}\label{tb:app}
\centering
    \begin{tabular}{|p{2.1cm}|p{1.7cm}|p{3.8cm}|}
    \hline
     \textbf{Application } & \textbf{Container} & \textbf{Applied optimization} \\\hline
     Monte Carlo & Array (small) & Vectorization friendly map \\\hline
     Black Scholes & N/A & Vectorization friendly map \\\hline
     Word Count & Hash table & SIMD hash, pipelining  \\\hline
     Reverse Index & Hash table & SIMD hash, pipelinine \\\hline
     Histogram & Array (large) & Eliminating local arrays \\\hline
     Bloom Filter & Array (large) & Eliminating local arrays \\\hline
    \end{tabular}
\end{table}\vspace{-0.3cm}

\textbf{Benchmark applications.} We choose six MapReduce applications as shown in Table \ref{tb:app}. Particularly, \textit{Histogram}, \textit{Word Count}, and \textit{Reverse Index} are the sample applications from Phoenix++. We implement \textit{Monte Carlo} and \textit{Black Scholes}, which follow the GPU-based parallel implementations \cite{cuda}. We also implement the building phase of \textit{Bloom Filter}, which simulates the use in bioinformatics \cite{bloomfilter}.

\textbf{Implementation detail.} MRPhi is developed using C++ and pthread. It natively runs on the Xeon Phi coprocessor. We organize the thread affinity in the scatter way such as thread $i$ belong to core $(i \mod 60)$, where 60 is the number of cores.

For the experiments, we first chracterize the performance of Xeon Phi coprocessor. These early characterization results motivate our design and are also useful for other developers. Then we study the performance impact of our various optimization techniques. Next, we conduct end-to-end performance comparison with Phoenix++ on the Xeon Phi. Finally, we report the performance comparison with a traditional Intel Xeon CPU. By default, the number of threads per core is set to the one that can generate the best performance, unless specified otherwise.
\vspace{-0.1cm}
\subsection{Characterizing Xeon Phi Coprocessor}
\vspace{-0.1cm}

\begin{figure}[ht]
\centering
\includegraphics[scale=0.5]{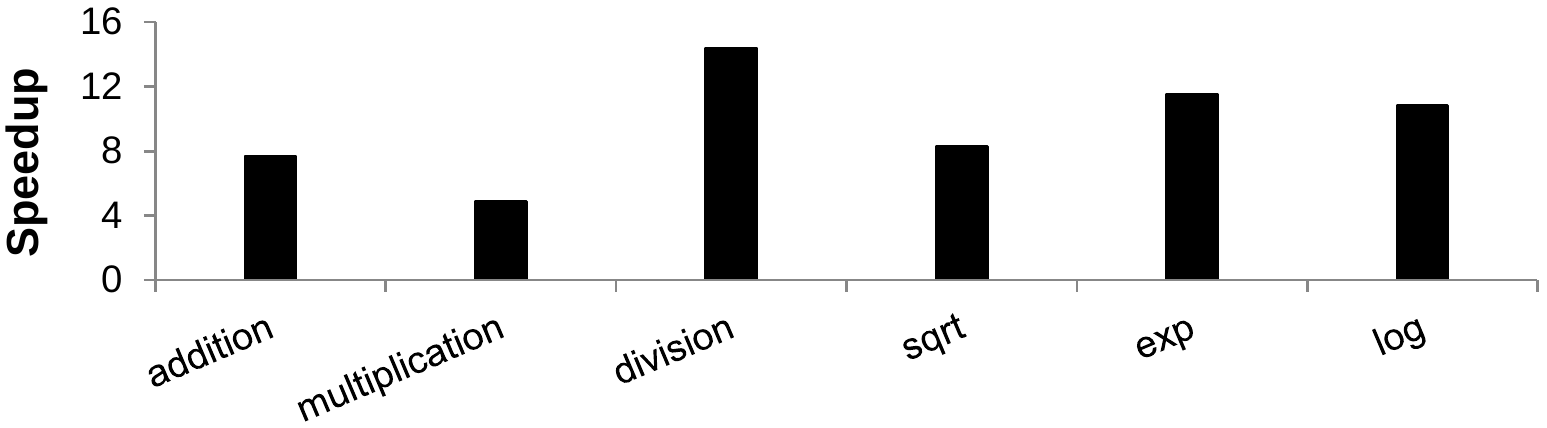}
\vspace*{-.4cm}
\caption{Speedup of vectorization for mathematical functions (a single thread).}
\label{fig:vectorization} \vspace*{-.4cm}
\end{figure}

\textbf{Auto-vectorization performance.} We first evaluate the performance of auto-vectorization for computation-intensive workloads. For each mathematical function, we use one thread to evaluate a large number of input elements. Figure \ref{fig:vectorization} reports the speedup of using vectorization over their scalar versions. It shows 5-14X speedup can be achieved employing vectorization. This confirms that utilizing VPUs is crucial for high performance.

\begin{figure}[ht]
\centering
\includegraphics[scale=0.52]{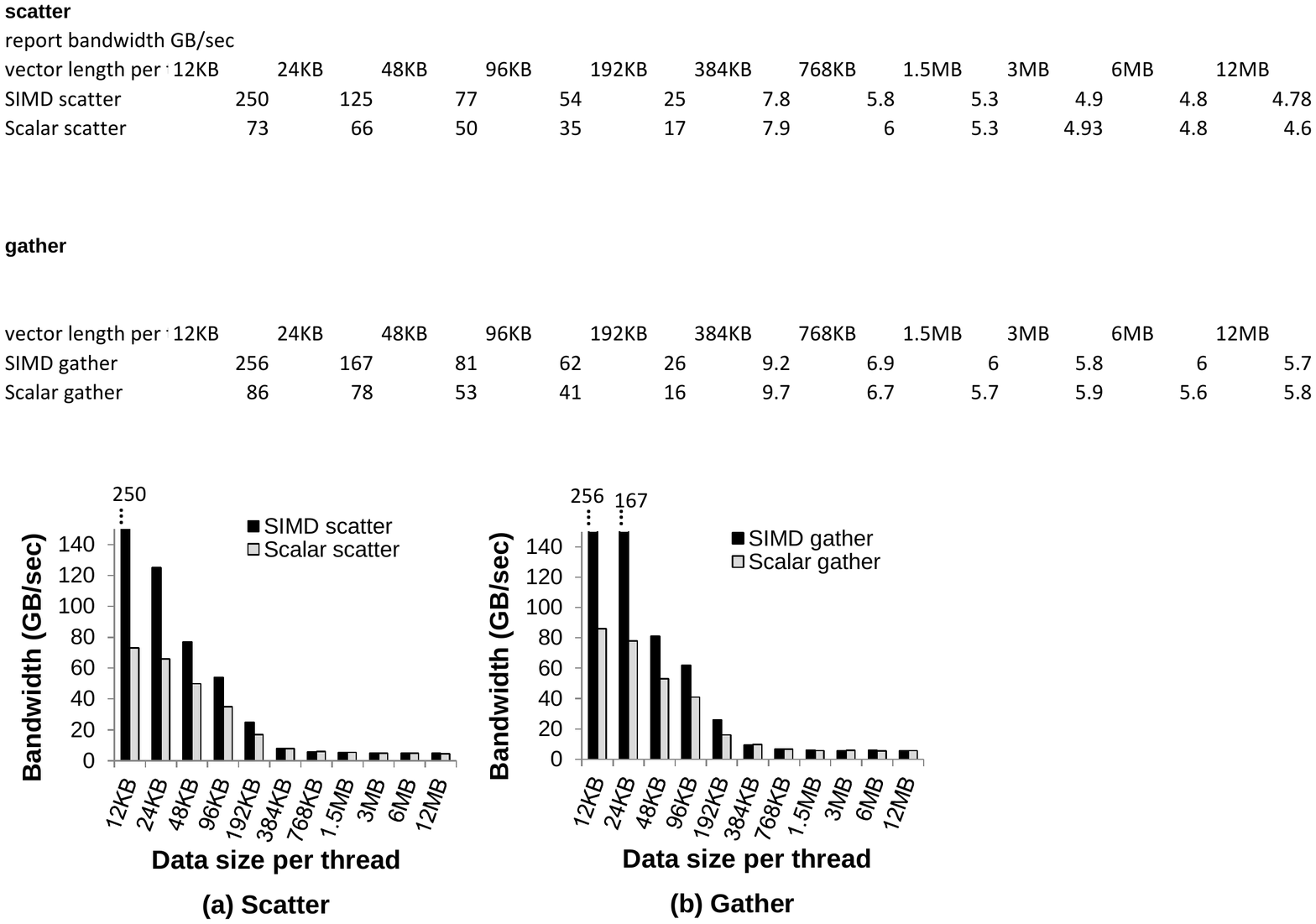}
\vspace*{-.4cm}
\caption{Comparison between SIMD and scaler scatter/gather.}
\label{fig:scatter_gather} \vspace*{-.5cm}
\end{figure}

\textbf{SIMD scatter and gather. } We evaluate the new SIMD scatter/gather instructions on the Xeon Phi. We make each thread perform independent scatter and gather (each thread has its own input and output arrays). We vary the total size of the data per thread to study the performance. Figure \ref{fig:scatter_gather} shows that when the data size is small to fit into the local L2 cache, SIMD scatter and gather are up to 3.4X faster than their scalar versions. However, when the data size becomes large, the SIMD scatter and gather do not help the performance since the performance is dominated by the memory latency due to cache misses. This suggests us that it is worthwhile to exploit the SIMD scatter/gather when the memory accesses are distributed in a small range of addresses. For example, it is efficient for SIMDH-Padding to use the SIMD scatter/gather to pack data to the vector registers.

\begin{figure}[ht]
\centering
\includegraphics[scale=0.50]{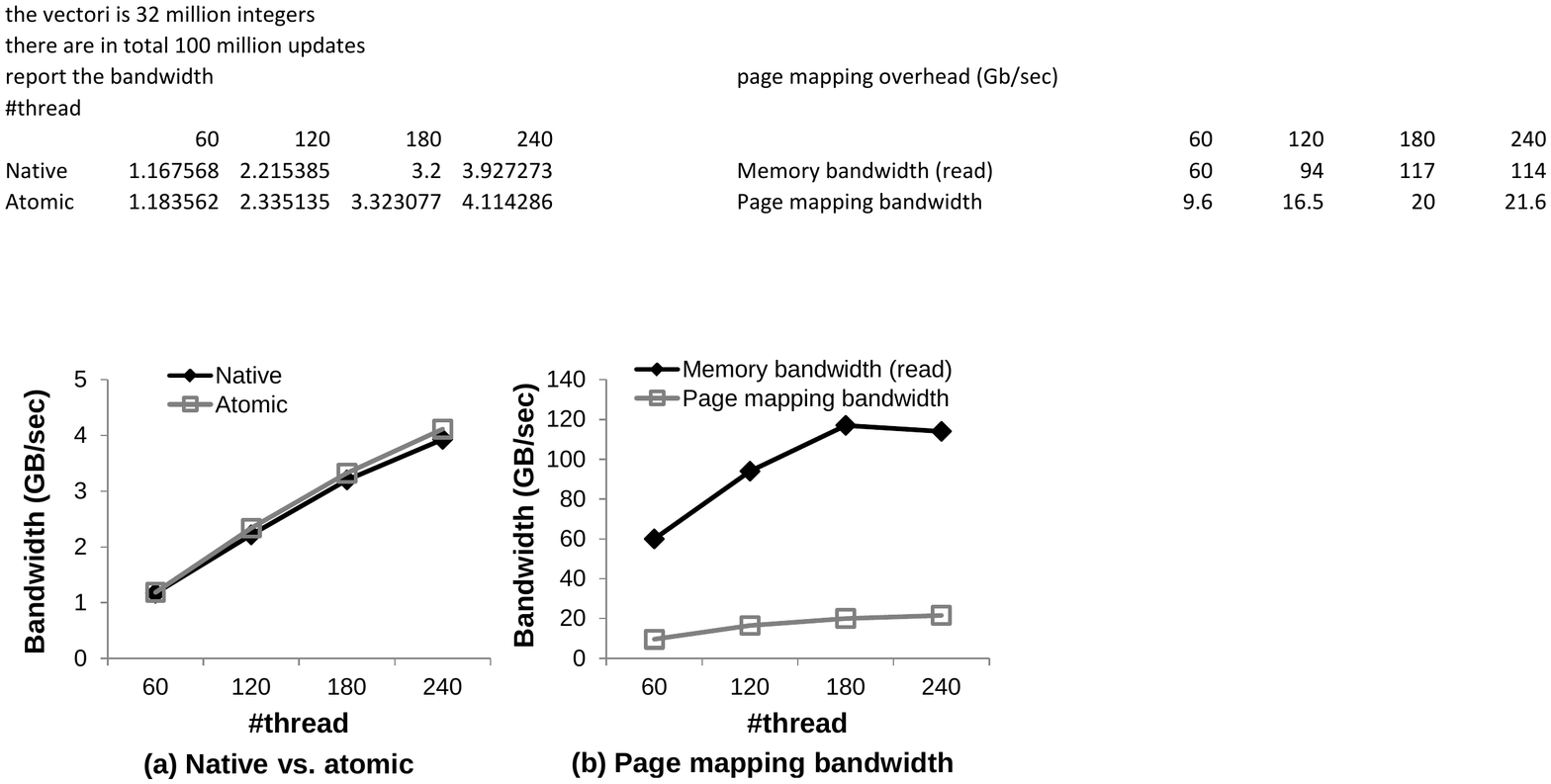}
\vspace*{-.3cm}
\caption{(a) The random memory access bandwidth with native and atomic data types. (b) Memory bandwidth of page mapping.}
\label{fig:atomic_init} \vspace*{-.2cm}
\end{figure}

\textbf{Atomic data types.} We study the performance of atomic operations for our purpose. Our context is random memory accesses on a large array with a very low conflict rate. We design our experiment to randomly update elements in an array with 32 million integers. Figure \ref{fig:atomic_init}(a) shows that using the native and atomic data types do not have much performance difference. Note that we guarantee the update with native data types does not have conflicts. We consider the overhead of atomic operations is hidden by the memory latency. This suggests us when the memory accesses are random with a low conflict rate, using atomic data types on the Xeon Phi is a reasonable choice when designing algorithms.
\vspace{-0.1cm}

\textbf{Thread initialization overhead.} Finally, we find the thread initialization overhead on the Xeon Phi is high. The initialization overhead on the Xeon Phi is around 0.75 millisecond per thread, while only around 0.067 millisecond per thread on the Xeon CPU. Therefore, we implement a thread pool and make the threads be initialized only once (with the total overhead of around 240 millisecond). In our evaluation, we exclude this thread pool initialization overhead since it is the same for all programs on the Xeon Phi.

\vspace{-0.1cm}
\subsection{Performance Evaluation of Optimization Techniques} \label{sec:exp_study}
\vspace{-0.1cm}
In this section, we evaluate the performance impact of our proposed techniques in detail. When we evaluate a specific technique, we evaluate the optimized implementation (with all applicable techniques enabled, denoted as {\em Opt}.) and the other implementation without this specific technique. Both of these two implementations are implemented by ourselves. By default, the number of threads per core is set to the one that can generate the best performance, unless specified otherwise.

\begin{figure}[ht]
\centering
\includegraphics[scale=0.52]{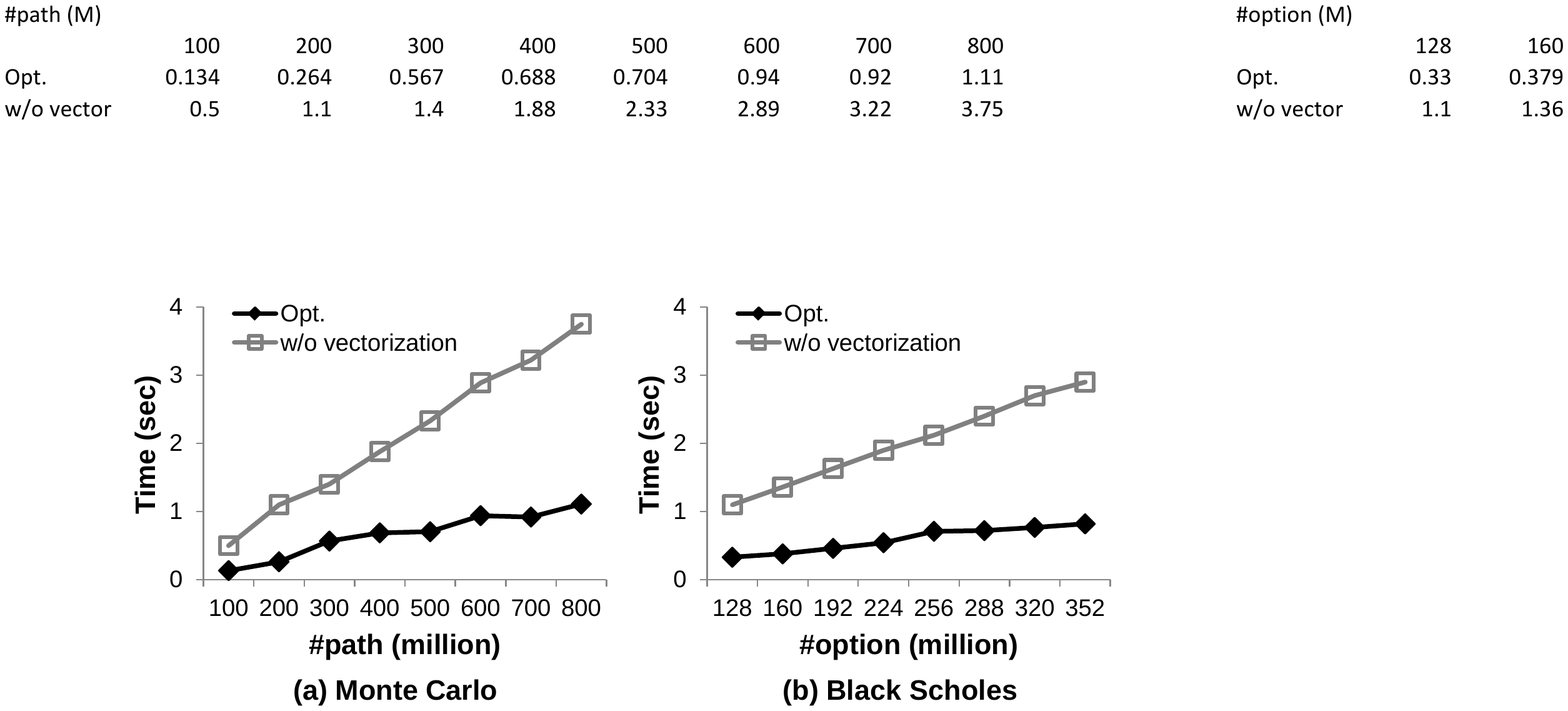}
\vspace*{-.4cm}
\caption{Performance impact of vectorization friendly map for Monte Carlo (with the number of paths varied) and Black Scholes (with the number of options varied).}
\label{fig:mc_bs} \vspace*{-.3cm}
\end{figure}

\textbf{Vectorization friendly map.} Figure \ref{fig:mc_bs} shows the performance impact of vectorization friendly map for {Monte Carlo} and {Black Scholes} with data size varied. It shows that the vectorization friendly map can improve the performance by 2.5-4.2X and 3.0-3.6X for Monte Carlo and Black Scholes, respectively. For those two applications, the map phase dominates the overall performance ($>$99\%). Therefore the vectorization for the map phase can greatly improve the overall performance.



\begin{figure}[ht]
\centering
\includegraphics[scale=0.52]{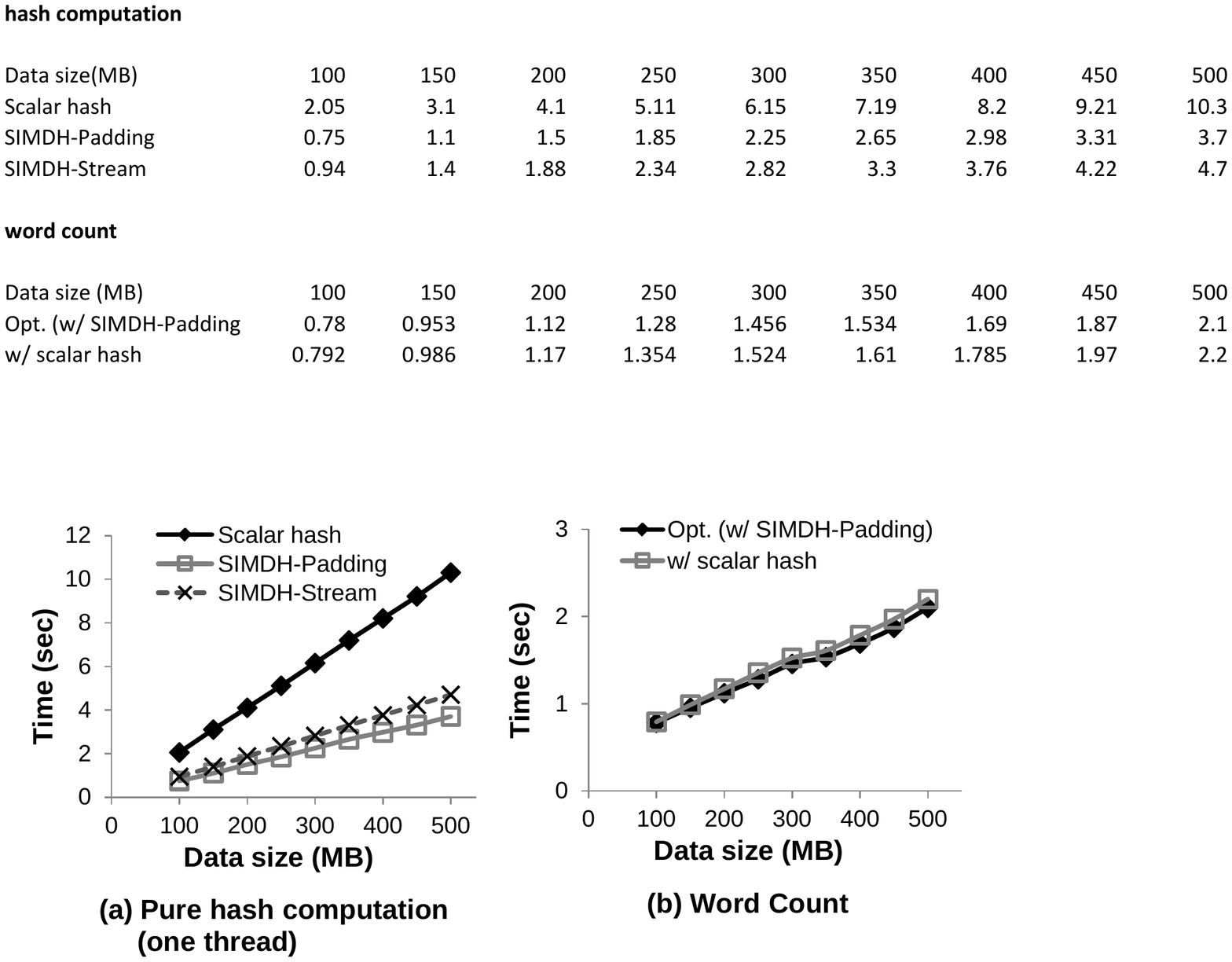}
\vspace*{-.3cm}
\caption{The SIMD hash computation performance with the input data size varied. (a) The pure hash computation time (one thread). (b) The performance of Word Count with and without the SIMD hash computation.}
\label{fig:hash} \vspace*{-.3cm}
\end{figure}

\textbf{SIMD parallelism for hash computation}. We first evaluate the performance of hash computation separately using a single thread. Figure \ref{fig:hash}(a) shows the performance result of pure hash computation with the input data size varied. We use the same data set as that used in the Word Count application. This shows that the SIMDH-Padding and SIMDH-Stream achieve the speedup of up to 2.8X and 2.2X over the scalar hash, respectively. Though SIMDH-Padding wastes computation resource due to the padding, it achieves better performance. There are two reasons. First, the SIMD scatter/gather operations are more efficient used in SIMDH-Padding due to the small address range of memory accesses. Second, the control logic of SIMH-Padding is simpler, thus the overhead of data padding to vectors is also lower.

We further show the performance impact of using the SIMD hash for Word Count in Figure \ref{fig:hash}(b). It shows the overall performance is slightly improved (around 6\%). The insignificant improvement is because the overall performance is dominated by the memory latency rather than the hash computation \cite{chen04}.

\begin{figure}[ht]
\centering
\includegraphics[scale=0.52]{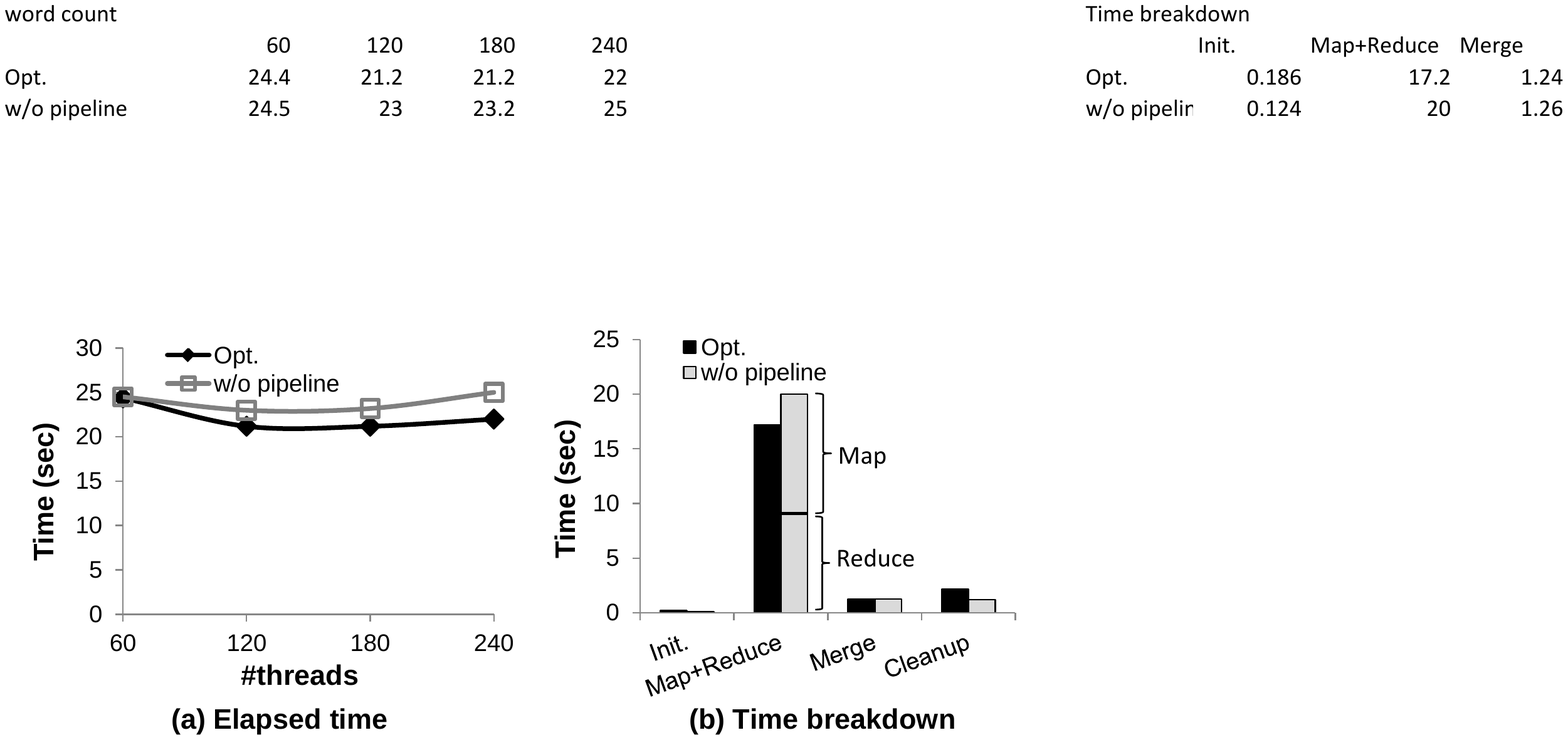}
\vspace*{-.3cm}
\caption{Performance impact of pipelined map and reduce for Reverse Index. (a) Elapsed time (b) Time breakdown.}
\label{fig:ri_pipeline} \vspace*{-.3cm}
\end{figure}

\textbf{Pipelined Map and Reduce.} Word Count and Reverse Index are able to take advantage of pipelined map and reduce. We report the results of Reverse Index as Word Count has the similar conclusion. We use the data set in Phoenix++ for evaluations, which contains 78,355 files and 307,921 links in total. Figure \ref{fig:ri_pipeline}(a) shows the elapsed time with the number of threads varied. It shows that the overall performance is improved by around 8.5\%. We further decompose the time as shown in Figure \ref{fig:ri_pipeline}(b). This shows that for the map and reduce phases only, the pipelining technique improves the performance by around 14\%. However, due to the storage overhead, the memory cleanup phase of the pipelined map and reduce is more expensive and offsets the overall performance improvement.

\textbf{Eliminating local arrays.} Now we study how the performance can be improved by eliminating local arrays but using atomic operations on the global array.

\begin{figure}[ht]
\centering
\includegraphics[scale=0.52]{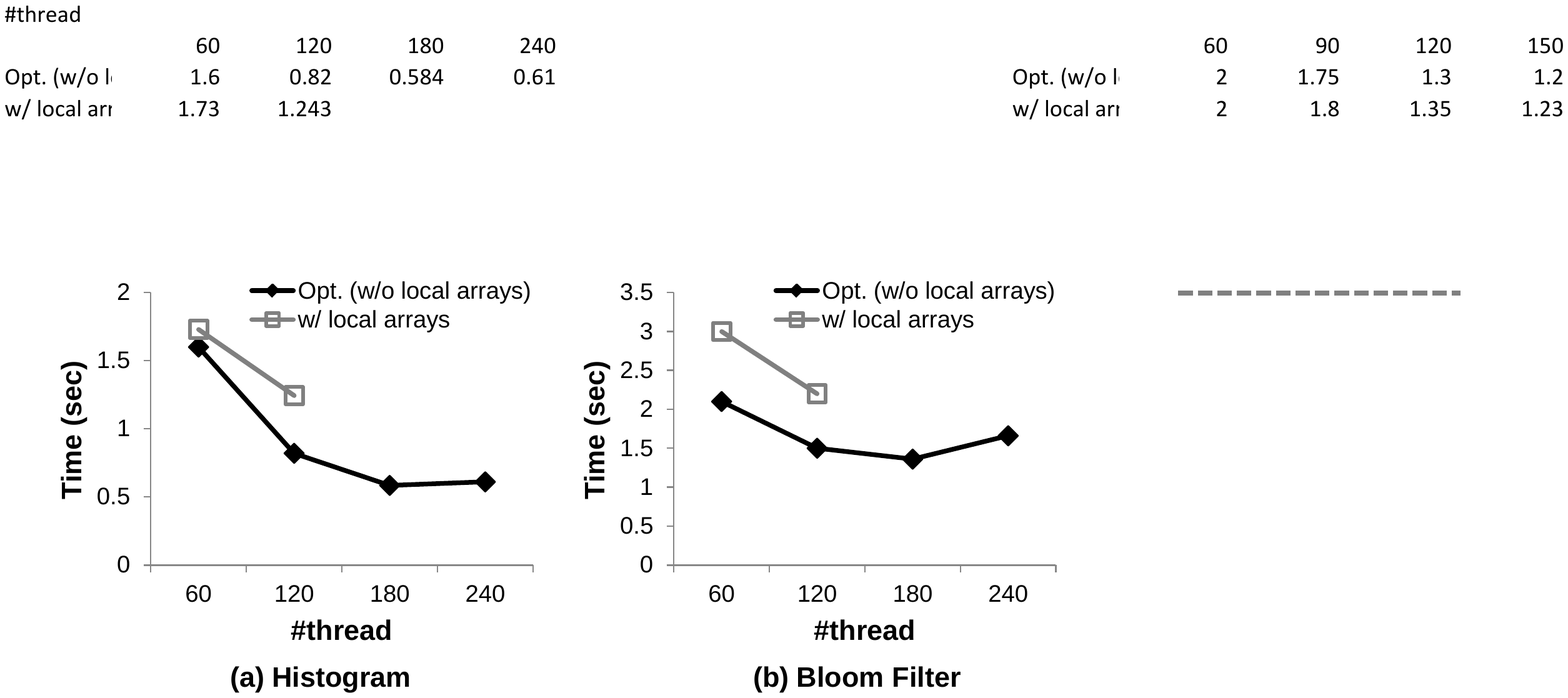}
\vspace*{-.4cm}
\caption{Elapsed time of Histogram and Bloom Filter with the number of map threads varied. Histogram: 16 million unique keys and 256 million input elements; Bloom Filter: 30 million input elements, 300 million entries.}
\label{fig:global_con} \vspace*{-.3cm}
\end{figure}

By eliminating local arrays, the thread scalability for the map phases can be improved. Figure \ref{fig:global_con} demonstrates such scenarios. The sizes of each local array are 64 MB and 40 MB for the Histogram and Bloom Filter, respectively. Figure \ref{fig:global_con} shows that the largest numbers of threads when using local arrays are 120 for Histogram and Bloom Filter, due to the limited memory size (8 GB). On the contrary, if local arrays are eliminated, more threads can be used. As a result, by eliminating local arrays, it achieves the speedup of up to 2.1X and 1.6X for Histogram and Bloom Filter, respectively. \mian{Note that, with the different sizes of arrays, the available number of map threads for using local arrays is different. Therefore, the performance improvement from eliminating local arrays varies acorrs different data sets.}

\begin{figure}[ht]
\centering
\includegraphics[scale=0.52]{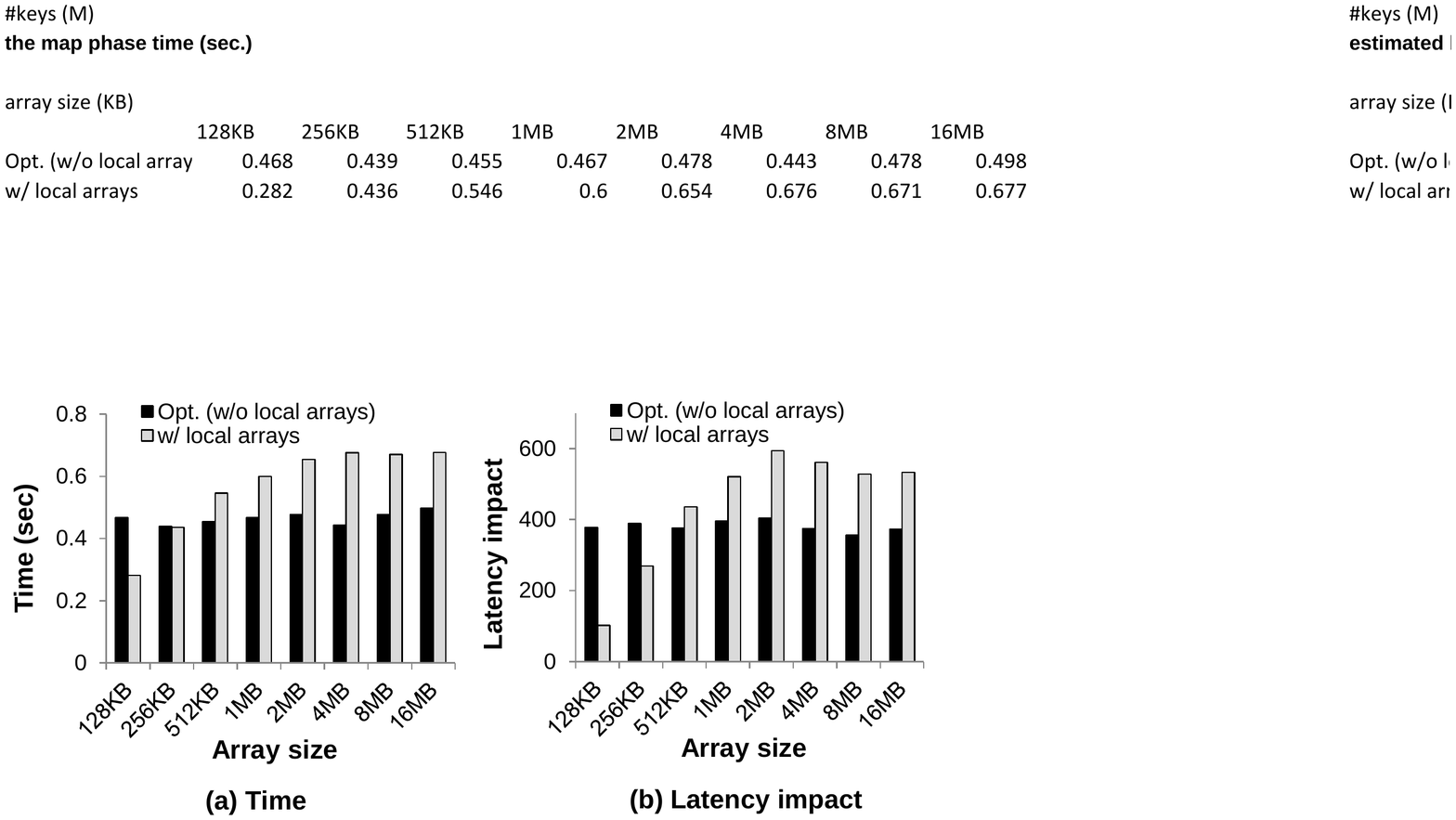}
\vspace*{-.4cm}
\caption{Performance impact of eliminating local arrays for the map phase in Histogram. (a) Time. (b) Estimated memory latency impact.}
\label{fig:global_locality} \vspace*{-.3cm}
\end{figure}

In Figure \ref{fig:global_con}, we also observe that when using the same number of threads, the optimized implementation still outperforms the implementation using local arrays. We consider this is because of the improved data locality. We further study this problem and report the results for Histogram. We vary the data size of each array (note that the Opt. solution only has one global array). In this experiment, we exclude the impact from the thread locality and make them be able to employ the same number of threads. Figure \ref{fig:global_locality}(a) shows that when the array size is small enough to fit into the L2 cache, using local arrays is more efficient. This is because the global array has the overhead of cache coherence. However, when the array becomes larger, using global arrays outperforms local arrays by up to 34\%. In such a case, both local and global arrays suffer from cache misses. However, the optimized solution can take advantage from the ring interconnection for better cache efficiency (Section \ref{sec:global}).

\mian{To confirm the cache efficiency, Figure \ref{fig:global_locality}(b) further shows the estimated memory latency impact. The memory latency impact is suggested by Intel to investigate the cache efficiency. It is an approximation of the number of clock cycles devoted to each L1 cache miss. Figure \ref{fig:global_locality}(b) shows the consistent trend of the memory latency impact as the elapsed time.}

\vspace{-0.1cm}
\subsection{MRPhi vs. Phoenix++ on the Xeon Phi}
\vspace{-0.1cm}

\begin{figure*}[ht]
\centering
\includegraphics[scale=1.24]{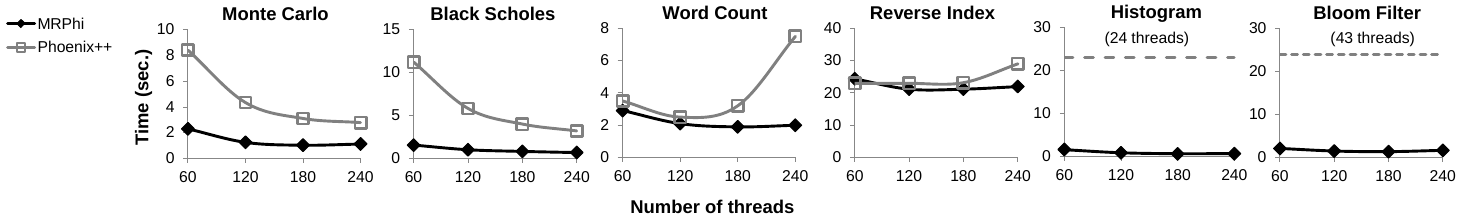}
\vspace*{-.4cm}
\caption{Performance comparison between MRPhi and Phoenix++ on the Xeon Phi. The horizontal axis is number of threads. The vertical axis is time (second).}
\label{fig:all} \vspace*{-.5cm}
\end{figure*}

\begin{table} [h]
  \caption{Data sets for end-to-end performance comparison.}\label{tb:data}
\centering
\vspace{-0.2cm}
    \begin{tabular}{|p{2.2cm}|p{5.8cm}|}
    \hline
     \textbf{Application } & \textbf{Data set} \\\hline
     Word Count & Input data size: 500 MB  \\\hline
     Reverse Index & \#files; 78,355 ; \#links: 307,921 ; size: 1 GB \\\hline
     Monte Carlo & \#paths: 800 million  \\\hline
     Black Scholes & \#options: 352 million  \\\hline
     Histogram & \#unique keys: 16 million; \#elements: 256 million   \\\hline
     Bloom Filter & \#elements: 30 million; \#entries: 300 million (Arabidopsis chromosome 1) \\\hline
    \end{tabular}\vspace{-0.3cm}
\end{table}

Now we show the end-to-end performance comparison between our MRPhi and Phoenix++ \cite{phoenix++}, which is state-of-the-art MapReduce framework on multicore. On the Xeon Phi, Phoenix++ also runs natively. We use large data sets for evaluations, which are summarized in Table \ref{tb:data}.

Figure \ref{fig:all} shows the overall performance comparison. Since the largest numbers of available threads for Histogram and Bloom Filter are less than 60 threads, we use dash lines. Overall, MRPhi outperforms Phoenix++ for all applications. For Monte Carlo and Black Scholes, which take advantage of vectorization in MRPhi, they are up to 2.7X and 4.6X faster than their counterparts in Phoenix++. For Word Count and Reverse Index that are based on the hash table, MRPhi can achieve the speedup of up to 1.2X. Furthermore, by eliminating local arrays, MRPhi is able to achieve the speedup of up to 38X and 18X for Histogram and Bloom Filter, respectively. Note that these two speedup numbers are even better than those reported in Figure \ref{fig:global_con}. This is mainly because Phoenix++ has some implementation issues when processing large arrays, which make it have worse thread scalability as well as much worse performance than our own implementation using local arrays.

In summary, our MRPhi can achieve the speedup of 1.2X to 38X over Phoenix++ for various applications on the Xeon Phi.

\vspace{-0.3cm}
\subsection{Performance Comparison between Xeon Phi and Xeon}
\vspace{-0.2cm}

\begin{figure}[ht]
\centering
\includegraphics[scale=0.9]{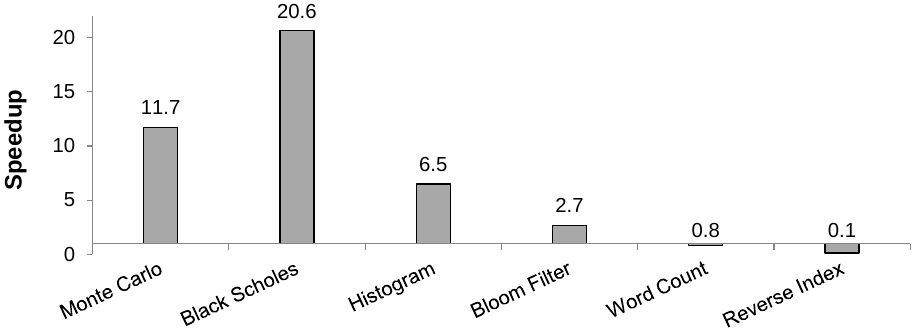}
\vspace*{-.4cm}
\caption{Speedup of MRPhi on Xeon Phi over Phoenix++ on Xeon. .}
\label{fig:vsxeon} \vspace*{-.3cm}
\end{figure}

Finally, we briefly compare the performance of MapReduce on Xeon Phi with that on a conventional Xeon processor. We adopt an Intel Xeon E5620 (2.4 GHz) CPU for comparison. The Xeon E5620 CPU has 4 cores and 8 threads in total. It has 12 MB cache. We report the performance of our MRPhi on the Xeon Phi and Phoenix++ on the Xeon. Since Phoenix++ is originally developed for multi-core CPUs. We use the data sets in Table \ref{tb:data}.

Figure \ref{fig:vsxeon} shows the performance comparison between the Xeon Phi and Xeon for six MapReduce applications. Particularly, for Monte Carlo and Black Scholes, which are able to employ vectorization to exploit the SIMD VPUs on the Xeon Phi, they can achieve the speedup of an order of magnitude on the Xeon Phi. This confirms that using vectorization or SIMD instructions is particularly important on the Xeon Phi.

On the other hand, Figure \ref{fig:vsxeon} also shows that Word Count and Reverse Index on the Xeon Phi are not as fast as those on the Xeon. From our further investigation, there are two major reasons. First, the hardware design of small local cache per core on the Xeon Phi is inefficient when there are a large number of random memory accesses, such as Word Count and Reverse Index using hash tables. Second, other system overhead on the Xeon Phi is significantly higher than that on the Xeon. From the performance tuning report of Intel VTune, we find the system overhead (functions from the Linux kernel or pthread library) takes 30-50\% of the overall Xeon Phi running time. On the Xeon, this percentage is lower than 10\%.


\vspace{-0.2cm}
\section{Conclusion} \label{sec:conclusion}
\vspace{-0.1cm}

In this work, we develop MRPhi, which is the first MapReduce framework optimized for the Intel Xeon Phi coprocessor. In MRPhi, in order to take advantage of VPUs, we develop a vectorization friendly technique for the map phase and SIMD hash computation. We also pipeline the map and reduce phases to better utilize the hardware resource. Furthermore, we eliminate local arrays to improve the thread scalability and data locality. Our framework is able to identify suitable techniques for a given application automatically. Our experimental results show that MRPhi can achieve the speedup of 1.2X to 38X over Phoenix++ for different applications.

\vspace{-0.15cm}
\bibliographystyle{IEEEtran}
\bibliography{main}

\begin{thebibliography}{10}
\providecommand{\url}[1]{#1}
\csname url@samestyle\endcsname
\providecommand{\newblock}{\relax}
\providecommand{\bibinfo}[2]{#2}
\providecommand{\BIBentrySTDinterwordspacing}{\spaceskip=0pt\relax}
\providecommand{\BIBentryALTinterwordstretchfactor}{4}
\providecommand{\BIBentryALTinterwordspacing}{\spaceskip=\fontdimen2\font plus
\BIBentryALTinterwordstretchfactor\fontdimen3\font minus
  \fontdimen4\font\relax}
\providecommand{\BIBforeignlanguage}[2]{{%
\expandafter\ifx\csname l@#1\endcsname\relax
\typeout{** WARNING: IEEEtran.bst: No hyphenation pattern has been}%
\typeout{** loaded for the language `#1'. Using the pattern for}%
\typeout{** the default language instead.}%
\else
\language=\csname l@#1\endcsname
\fi
#2}}
\providecommand{\BIBdecl}{\relax}
\BIBdecl

\bibitem{Bingshengsigmod}
B.~He, K.~Yang, R.~Fang, M.~Lu, N.~Govindaraju, Q.~Luo, and P.~Sander,
  ``Relational joins on graphics processors,'' in \emph{SIGMOD}, 2008.

\bibitem{gpustream}
H.~P. Huynh, A.~Hagiescu, W.-F. Wong, and R.~S.~M. Goh, ``Scalable framework
  for mapping streaming applications onto multi-gpu systems,'' in \emph{PPoPP},
  2012.

\bibitem{fpgamc}
J.~Castillo, J.~Bosque, E.~Castillo, P.~Huerta, and J.~Martinez, ``Hardware
  accelerated montecarlo financial simulation over low cost fpga cluster,'' in
  \emph{IPDPS}, 2009.

\bibitem{cuda}
``{NVIDIA Cuompute Unified Device Architecture (CUDA)},''
  \url{http://www.nvidia.com/object/cuda_home_new.html}.

\bibitem{opencl}
``{The open standard for parallel programming of heterogeneous systems},''
  \url{http://www.khronos.org/opencl/}.

\bibitem{verilog}
``Ieee standard verilog hardware description language,'' \emph{IEEE Std
  1364-2001}, 2001.

\bibitem{xeonphi1}
S.~Pennycook, C.~Hughes, M.~Smelyanskiy, and S.~Jarvis, ``Exploring simd for
  molecular dynamics, using intel xeon processors and intel xeon phi
  coprocessors,'' in \emph{IPDPS}, 2013.

\bibitem{xeonphi2}
A.~Heinecke, K.~Vaidyanathan, M.~Smelyanskiy, A.~Kobotov, R.~Dubtsov, G.~Henry,
  G.~Chrysos, and P.~Dubey, ``Design and implementation of the linpack
  benchmark for single and multi-node systems based on intel xeon phi
  coprocessor,'' in \emph{IPDPS}, 2013.

\bibitem{stampede}
``{STAMPEDE: Dell PowerEdge C8220 culster with Intel Xeon Phi Coprocessors},''
  \url{http://www.tacc.utexas.edu/resources/hpc/stampede}.

\bibitem{tianhe2}
``{China's Tianhe-2 Supercomputer Takes No. 1 Ranking on 41st TOP500 List},''
  \url{http://www.top500.org/blog/lists/2013/06/press-release/}.

\bibitem{google}
J.~Dean and S.~Ghemawat, ``Mapreduce: Simplified data processing on large
  clusters,'' in \emph{OSDI}, 2004.

\bibitem{phoenix}
C.~Ranger, R.~Raghuraman, A.~Penmetsa, G.~Bradski, and C.~Kozyrakis,
  ``Evaluating mapreduce for multi-core and multiprocessor systems,'' in
  \emph{HPCA}, 2007.

\bibitem{mars}
B.~He, W.~Fang, Q.~Luo, N.~K. Govindaraju, and T.~Wang, ``Mars: a mapreduce
  framework on graphics processors,'' in \emph{PACT}, 2008.

\bibitem{phoenix++}
J.~Talbot, R.~M. Yoo, and C.~Kozyrakis, ``Phoenix++: modular mapreduce for
  shared-memory systems,'' in \emph{Proceedings of the second international
  workshop on MapReduce and its applications}, 2011.

\bibitem{tiled}
R.~Chen, H.~Chen, and B.~Zang, ``Tiled-mapreduce: optimizing resource usages of
  data-parallel applications on multicore with tiling,'' in \emph{PACT}, 2010.

\bibitem{metis}
Y.~Mao, R.~Morris, and M.~F. Kaashoek, ``Optimizing mapreduce for multicore
  architectures,'' \emph{Computer Science and Artificial Intelligence
  Laboratory, Massachusetts Institute of Technology, Tech. Rep}, 2010.

\bibitem{multigpu}
J.~A. Stuart and J.~D. Owens, ``Multi-gpu mapreduce on gpu clusters,'' in
  \emph{IPDPS}, 2011.

\bibitem{mapcg}
C.~Hong, D.~Chen, W.~Chen, W.~Zheng, and H.~Lin, ``Mapcg: writing parallel
  program portable between cpu and gpu,'' in \emph{PACT}, 2010.

\bibitem{apu}
L.~Chen, X.~Huo, and G.~Agrawal, ``Accelerating mapreduce on a coupled cpu-gpu
  architecture,'' in \emph{Supercomputing}, 2012.

\bibitem{fpga}
Y.~Shan, B.~Wang, J.~Yan, Y.~Wang, N.~Xu, and H.~Yang, ``Fpmr: Mapreduce
  framework on fpga,'' in \emph{FPGA}, 2010.

\bibitem{cell}
M.~de~Kruijf and K.~Sankaralingam, ``Mapreduce for the cell broadband engine
  architecture,'' \emph{IBM J. Res. Dev.}, vol.~53, no.~5, Sep. 2009.

\bibitem{fnv}
``{FNV Hash},'' \url{http://www.isthe.com/chongo/tech/comp/fnv/index.html}.

\bibitem{djb2}
``{Hash Functions},'' \url{http://www.cse.yorku.ca/~oz/hash.html}.

\bibitem{online}
T.~Condie, N.~Conway, P.~Alvaro, J.~M. Hellerstein, K.~Elmeleegy, and R.~Sears,
  ``Mapreduce online,'' in \emph{NSDI}, 2010.

\bibitem{bloomfilter}
L.~Ma, R.~D. Chamberlain, J.~D. Buhler, and M.~A. Franklin, ``Bloom filter
  performance on graphics engines,'' in \emph{ICPP}, 2011.

\bibitem{chen04}
S.~Chen, A.~Ailamaki, P.~Gibbons, and T.~Mowry, ``Improving hash join
  performance through prefetching,'' in \emph{Data Engineering, 2004.
  Proceedings. 20th International Conference on}, 2004.

\end{thebibliography}

\end{document}